\documentstyle[12pt]{article}

\def\be{\begin{equation}}
\def\ee{\end{equation}}
\def\ba{\begin{array}{c}}
\def\ea{\end{array}}
\def\p{\partial}
\def\ben{$$}
\def\een{$$}
\begin{document}

\titlepage
%\vspace*{2cm}

\begin{center}{\Large \bf
A generalization of the concept of ${\cal PT}$ symmetry
 }\end{center}

\vspace{5mm}

\begin{center}

{\bf Miloslav Znojil} \vspace{3mm}

Theory Group,
Nuclear Physics Institute AS CR,
\\CS 250 68 \v{R}e\v{z},
Czech Republic\footnote{e-mail: znojil@ujf.cas.cz
 % \small \today, genp.tex file, J. Phys. A
 }

\vspace{5mm}

\end{center}

\section*{Abstract}
We show that the ${\cal PT}$ symmetric Hamiltonians (and their
generalizations $H=H^\ddagger$ defined in the text) may be all
assigned the projected (so called Feshbach or effective)
nonlinear Hamiltonians which are ``locally" Hermitian. This
implies that many (if not all) of the bound-state energies may
be real in a broad domain of Hermiticity-violating interactions.
A complexification of a superintegrable $D=2$ example is
conjectured as an illustration.

 \vspace{9mm}

\noindent PACS 03.65.Bz, 03.65.Ca 03.65.Fd

%\begin{center}
%\end{center}

\newpage

\section{Introduction}

 % \noindent
Evolution of bound states in quantum mechanics is mediated (or
generated) by their Hamiltonian,  $
 |\psi(t)\rangle = \exp (-i\,H\,t)\,
 |\psi(0)\rangle$.
In the models with Hermitian $H =H^\dagger$ the availability of
solutions of the time-independent Schr\"{o}dinger equation
 \be
 H\,|\psi_n\rangle =
 E_n\,|\psi_n\rangle, \ \ \ \ \ \ n = 0, 1, \ldots
 \label{SE}
 \ee
simplifies this rule since all the eigenvalues $E_n$ remain real
and the time-dependence of the separate eigenstates becomes
elementary,
 \be
 |\psi_n(t)\rangle = e^{-i\,E_n\,t}\,
 |\psi_n(0)\rangle\ .
 \label{dva}
 \ee
A puzzling situation is encountered when the real energies $E_n$
are derived from a non-Hermitian Hamiltonian $H \neq H^\dagger$.
Recently, Bender et al \cite{BB} conjectured that many models of
such a type exist and are characterized by a ``weaker
analogue" of Hermiticity, $H = H^\ddagger$. For the sake of
definitness they restricted their attention to a small class of
the complex one-dimensional models and performed a number of
numerical and semi-classical calculations showing that the spectra
$\{E_n\}$ of their non-Hermitian $ H=H^\ddagger =
p^2+x^{2N}(ix)^\epsilon$ with $\epsilon>0$ are real,
discrete and bounded below. Generalizing this example
they conjectured the definition
 \be
 H^\ddagger = {\cal PT}\,H\,{\cal PT}
 \label{PT}
 \ee
of the required ``weaker Hermiticity". The operator ${\cal P}$ is
defined as changing the parity, ${\cal P}\, x = -x$, while ${\cal
T}$ mimics the time reversal, ${\cal T}\,i=-i$. Our present
remark is inspired by the comparatively narrow  range of the
existing applications of the definition~(\ref{PT}) which are
mainly single-particle or one-dimensional
(cf.~\cite{a}-\cite{mytri}).

In a preparatory step we shall clarify an algebraic background
of the apparently unmotivated assumption (\ref{PT}) (section 2).
In the main body of this paper (sections 3 and 4) we shall
propose a more general definition of the weakened Hermiticity
$H=H^\ddagger$. An applicability of the scheme is illustrated on
a complexification of an elementary two-dimensional
superintegrable example of ref.~\cite{Winternitz}.

Our definition extends the ${\cal PT}$ symmetric class of
non-Hermitian Hamiltonians supporting the real spectra and
generating the oscillatory, unitary-like time-dependence
(\ref{dva}) of bound states in quantum mechanics.  In
section 5 we shall argue that the generalized scheme
(with the parity replaced by a more general operator ${\cal Q}$)
remains mathematically selfconsistent. It admits many physical
interpretations of the operator $ {\cal Q}$ which plays the role
of an indefinite metric in our Hilbert (or rather Krein or
Pontrjagin) space of admissible
wavefunctions~\cite{pseudo,pseudob,ioch}.

\section{Explanation}

The ${\cal PT}$ symmetric Hamiltonians $ H = {\cal PT}\,H\,{\cal
PT}$ (with ${\cal P}^2=1$ and ${\cal T}^2=1$) which possess a
real-energy solution $|\psi_n\rangle$ of eq. (\ref{SE}) resemble
their Hermitian analogues in several aspects.
They may be split in the real and imaginary part, $H =
S + i\,A$ and, since $  {\cal P}\,H\,{\cal P} = {\cal T}\,H\,{\cal
T}$ by assumption, we have $  {\cal
P}\,S\,{\cal P} = S$, $  {\cal P}\,A\,{\cal P} = -A$.
Each wavefunction $ |\psi_n\rangle$ may be complemented
by another eigenstate $ {\cal PT}\,|\psi_n\rangle$ at the same
(real) energy $E_n$.  In the generic non-degenerate case this
means that among the two linearly dependent superposition
solutions $  \pm |\psi_n\rangle+{\cal PT}\,|\psi_n\rangle =
|\psi^{[\pm]}_n\rangle$ of eq. (\ref{SE}), we are free to pick up
one with the even or odd ${\cal PT}-$parity.
In what
follows we shall assume that the latter generalized parity has
been fixed as positive which means that we can
write
 \ben
 |\psi_n\rangle = | \sigma_n \rangle + i\, |\tau_n\rangle,
\ \ \ \ \ \ \ \ \ \
  {\cal P}\,| \sigma_n \rangle = | \sigma_n \rangle,
 \ \ \ \ \ \ \ \ \ \
  {\cal P}\, |\tau_n\rangle
 = - |\tau_n\rangle\ .
  \een
In any (e.g., harmonic-oscillator)  basis $\{ |
n^{(\alpha)} \rangle \}$ numbered by the integers $n = 0, 1,
\ldots$ and by the even and odd parity $\alpha = \pm 1$ we can
expand
 \ben
  | \sigma_n \rangle=\sum_{k=0}^{\infty}\, | n^{(+)}
\rangle\, s_k,
 \ \ \ \ \ \ \ \ \ \
  |\tau_n\rangle
 =\sum_{n=0}^{\infty}\, | n^{(-)}
\rangle\, t_k
  \een
and re-write our Schr\"{o}dinger equation (\ref{SE}) in the linear
algebraic form
 \ben
  \sum_{k=0}^{\infty}\
  \langle m^{(+)}|\,S\,|\,k^{(+)}\rangle\
  s_k
  -
  \sum_{j=0}^{\infty}\
  \langle m^{(+)}|\,A\,|\,j^{(-)}\rangle\
  t_j
  =E\
  s_m,
   \een
  \ben
  \sum_{k=0}^{\infty}\
  \langle m^{(-)}|\,S\,|\,k^{(-)}\rangle\
  t_k
  +
  \sum_{j=0}^{\infty}\
  \langle m^{(-)}|\,A\,|\,j^{(+)}\rangle\
  s_j
  =E\
  t_m
 \label{SEgat}
  \een
with $ m = 0, 1, \ldots$. Switching to a compactified notation
  \ben
  \sum_{k=0}^{\infty}\
  S^{(+)}_{mk}
\,   s_k
  -
  \sum_{j=0}^{\infty}\
  A_{mj}\,
  t_j
  =E\
  s_m,
   \een
   \ben
  \sum_{k=0}^{\infty}\
S^{(-)}_{mk}\,
  t_k
  +
  \sum_{j=0}^{\infty}\
A_{mj}^\dagger\,
  s_j
  =E\
  t_m
 \label{iSEgat}
  \een
and to its further matrix (non-Hermitian) abbreviation
 \be
\left (
\begin{array}{cc}
F-E\,I&-A\\
A^\dagger&
G-E\,I
 \ea
\right )
\cdot \left (
\ba
\vec{s}\\
\vec{t}
 \ea
\right ) = 0\ ,
\label{SEd}
 \ee
we eliminate $\vec{t}=- \left ( G-E\,I \right ) ^{-1}A^\dagger
\vec{s}$ and get the reduced Schr\"{o}dinger equation
with the so called Feshbach or effective energy-dependent
Hamiltonian~\cite{Feshbach},
 \be
H_{eff}(E)
\vec{s}=E\,
\vec{s}, \ \ \ \ \ \
H_{eff}(E)
=
F + A\,
\left (
G-E\,I
\right )^{-1}
A^\dagger\ .
\label{Feshbach}
 \ee
We can formulate our first important conclusion: The reality of
the spectrum of many ${\cal PT}$ symmetric Hamiltonians is the
consequence of an elementary observation that their
Feshbach's effective Hamiltonian $H_{eff}(E)$ in eq.
(\ref{Feshbach}) can be approximated by its
energy-independent forms $H_{eff}(\varrho)$. All of these
effective Hamiltonians are Hermitian and possess the real spectra
$\{E_n(\varrho)\}$. The exact energy levels are obtained from them via
the nonlinear selfconsistency condition
 \be
\varrho = \varrho_n = E_n(\varrho).
\label{selfconsistency}
 \ee
For comparison, it is extremely useful to imagine that our
equation (\ref{SEd}) is formally related to the problem
where one replaces the upper right submatrix $-A$ by the
block $+A$ with an opposite sign. The new Hamiltonian becomes
Hermitian (and real and symmetric) and we
derive its current effective $H_{Herm.-eff}(E)$ which differs from
its ${\cal PT}$ symmetric predecessor in eq. (\ref{Feshbach})
by the artificial sign-change,
 \ben
H_{Herm.-eff}(E)
=
F - A\,
\left (
G-E\,I
\right )
^{-1}
A^\dagger.
 \een
Besides the Hermitian case, equation (\ref{selfconsistency}) may
have solely real solutions in a broad domain of the coupling
strengths in the ${\cal PT}$ symmetric regime. An elementary
illustration of such an expectation is provided by the
two-by-two matrix with the four real matrix elements,
 \ben
\left (
\begin{array}{cc}
f-E&-a\\
a&g-E
 \ea
\right )
\left (
\ba
{s}\\
{t}
 \ea
\right ) = 0\ .
 \een
This mimics our Schr\"{o}dinger eq. (\ref{SEd}) and the dimension
of its effective Hamiltonian is one. The exact spectrum
 \ben
E=E_{1,2}=\frac{1}{2}
\left (
f+g\pm \sqrt{
(f-g)^2-4a^2
}
\right )
 \een
proves all real and non-degenerate if and only if $2|a| < |f-g|$.
This is to be compared with the parallel Hermitian
illustration the spectrum of
which is {\em always} real,
 \ben
E_{Herm.-eff}=\frac{1}{2}
\left (
f+g\pm \sqrt{
(f-g)^2+4a^2
}
\right )\ .
 \een
We may summarize: In contrast to the Hermitian case the complete
reality of the spectrum of non-Hermitian models is not robust and
can be violated by a change of the magnitude of matrix elements.
The ${\cal PT}$ symmetry offers a firm guidance of our
understanding of the stability of the spectrum in terms of the
fairly transparent selfconsistency condition
(\ref{selfconsistency}). The frequent occurrence of the ${\cal
PT}$ symmetric models with real spectra has been constructively
confirmed by the numerous examples which are solvable
non-numerically, without any recourse to their matrix
representation (cf. refs.~\cite{b}).

\section{Generalization}

A core of our preceding explanation (why the ${\cal PT}$ symmetric
Hamiltonians (\ref{PT}) can have real energies) lies in the
demonstration of the Hermiticity of their effective form
$H_{eff}(\varrho)$ (so that all the auxiliary
$E_n(\varrho)$ are real). The Hermitian and
non-Hermitian alternative mechanisms of breaking the parity mean
that we start from a doublet of independent
even-parity and odd-parity Hamiltonians $F$ and $G$
and couple them in Schr\"{o}dinger equation
 \be
\left (
\begin{array}{cc}
F-E\,I&\alpha\,A\\
A^\dagger&
G-E\,I
 \ea
\right )
\cdot \left (
\ba
\vec{s}\\
\vec{t}
 \ea
\right ) = 0\
\label{SEdrd}
 \ee
where either $\alpha=1$ (Hermitian case) or $\alpha=-1$
(${\cal PT}$ symmetric case).
The partitioning need not necessarily be related to the usual
parity of the basis. Our argument has been entirely general. For
example, in the Feshbach's re-interpretation of eq.
(\ref{SEdrd}), the upper partition of the size ${\rm dim}\,F$
represents the more relevant part of
the Hilbert space determined by
the so called ``model space" projector.  The lower Feshbach's partition
is usually treated in less detail.  Thus, after we truncate the
basis in the Hilbert space (say, in a variational setting), we
can have very different partitions, with $m={\rm dim}\,F \neq
n={\rm dim}\,G$. This is a comparatively easy
generalization of the ${\cal PT}$ symmetry but does not seem to
exhaust all the possibilities.

\subsection{Partitioning three by three}

The next step of our analysis is based on the triple
partitioning of the basis, finite or
infinite. Let us assume that $\alpha, \beta, \gamma = \pm 1$ and
postulate a parallel between the ${\cal PT}$
symmetry and Hermiticity in the triply partitioned equation
 \ben
\left (
\begin{array}{ccc}
F-E\,I&\alpha\,A&
\beta\,B
\\
A^\dagger&
G-E\,I&\gamma\,C\\
B^\dagger&
C^\dagger&
Z-E\,I
 \ea
\right )
\cdot \left (
\ba
\vec{r}\\
\vec{s}\\
\vec{t}
 \ea
\right ) = 0\ .
\label{SEtre}
 \een
The
elimination of
 $
\vec{t}
=
-
\left (
Z-E\,I
\right )
^{-1}
\left (
B^\dagger
\vec{r}
+
C^\dagger
\vec{s}
\right )
 $
gives us the two by two effective
 Schr\"{o}dinger equation
 \be
\left [
\left (
\begin{array}{cc}
F-E\,I&\alpha\,A\\
A^\dagger&
G-E\,I
 \ea
\right )
-
\left (
\begin{array}{c}
\beta\,B\\
\gamma\,C
 \ea
\right )
\begin{array}{cc}
\left (
Z-\varrho\,I
\right )^{-1}
(B^\dagger&
C^\dagger)\\
&
 \ea
\right ]
\cdot \left (
\ba
\vec{r}\\
\vec{s}
 \ea
\right ) = 0
\label{FEdrd}
 \ee
where $ \varrho \equiv E$. We intend to
guarantee that the effective
Hamiltonians
$H_{eff}(\varrho)$
remain Hermitian.
At any $\varrho = constant$ the reality of all the energy roots
$E_n(\varrho)$ of the linearized eq. (\ref{FEdrd}) near $\varrho$
will be guaranteed
by this Hermiticity. It is true for the diagonal blocks in
$H_{eff}(\varrho)$.
In order to satisfy also the ``off-diagonal" condition,
 \ben
\alpha'\,A' = \alpha\,A -\beta\,B\,C^\dagger, \ \ \ \ \ \left(
{A}' \right )^\dagger = A^\dagger -\gamma\,C\,B^\dagger
 \een
we choose $\alpha' = \alpha$ and arrive at the
constraint
 \be
\alpha\beta=\gamma \label{rules}
 \ee
and menu
  \be
\begin{array}{|ccc|}
\hline
\alpha&\beta&\gamma\\
\hline
+&+&+\\
+&-&-\\
-&+&-\\
-&-&+\\
\hline
 \ea
 \label{tabl}
 \ee
which lists all the available possibilities. The first line
represents the Hermitian choice.

\subsection{Partitioning four by four}

Let us now preserve the latter rule (\ref{rules}), add the three
new variables $\mu, \nu, \rho = \pm 1$ and postulate
 \ben
\left (
\begin{array}{cccc}
F-E\,I&\alpha\,A&
\beta\,B&\mu\,U
\\
A^\dagger&
G-E\,I&\gamma\,C&\nu\,V\\
B^\dagger&
C^\dagger&
Z-E\,I&\rho\,W\\
U^\dagger&
V^\dagger&
W^\dagger&
K-E\,I
 \ea
\right )
\cdot \left (
\ba
\vec{p}\\
\vec{r}\\
\vec{s}\\
\vec{t}
 \ea
\right ) = 0\ .
\label{SEctere}
 \een
The insertion
of
$
\vec{t}
=
-
\left (
K-E\,I
\right )
^{-1}
\left (
U^\dagger
\vec{p}
+
V^\dagger
\vec{r}
+
W^\dagger
\vec{s}
\right )
 $
reduces the Schr\"{o}dinger equation to a three by three
partitioned problem
 \ben
\left [
\left (
\begin{array}{ccc}
F-E\,I&\alpha\,A&\beta\,B\\
A^\dagger&
G-E\,I&\gamma\,C\\
B^\dagger&
C^\dagger&
Z-E\,I
 \ea
\right )
\right .
 \een
 \ben
\left . -
\left (
\begin{array}{c}
\mu\,U\\ \nu\,V\\ \rho\,W  \ea \right )
\begin{array}{ccc} \left ( K-\varrho\,I \right
)^{-1}
(U^\dagger& V^\dagger& W^\dagger)\\ &&\\ &&  \ea  \right ] \cdot
\left ( \ba \vec{p}\\ \vec{r}\\ \vec{s}  \ea \right ) = 0, \ \ \ \
\ \ \varrho \equiv E\ . \label{FEhdrd}
  \een
In order that the effective Hamiltonians $H_{eff}(\varrho)$
preserve the three by three symmetry
$H=H^\ddagger$, we satisfy the elementary
``diagonal" conditions by fixing $\alpha'=\alpha,\ \beta'=\beta$ and
$\gamma'=\gamma$. The three ``off-diagonal" constraints
 \ben
\alpha\beta=\gamma,\ \ \ \ \ \ \ \ \alpha\mu=\nu,\ \ \ \ \ \ \ \
\beta\mu=\rho,\ \ \ \ \ \ \ \ \gamma\nu=\rho.
 \een
lead to the following eight solutions,
 \be
\begin{array}{|cccccc|}
\hline
\alpha&\beta&\gamma&\mu&\nu&\rho\\
\hline
+&+&+&+&+&+\\
+&+&+&-&-&-\\
+&-&-&+&+&-\\
+&-&-&-&-&+\\
-&+&-&+&-&+\\
-&+&-&-&+&-\\
-&-&+&-&+&+\\
-&-&+&+&-&-\\
\hline
 \ea
\label{tabulka}
 \ee
The first line is the Hermitian choice and the fourth line
reproduces the result of the two by two partitioning. The
non-square two by two partitioning gives the lines number two and
seven. The remaining four items offer the genuine three by three
structures. Two contain the three minuses and the other two the
four ones.

\subsection{Example}

The partitioning of the bases appears in the majority of their
(e.g., variational) applications. For illustration, let us
consider a two-dimensional Hamiltonian
 \ben
H = -\p_x^2-\p^2_y+x^2+y^2
+\frac{g}{x^2}
+\frac{g}{y^2}
 \een
which is superintegrable \cite{Winternitz}. As a consequence, its
two-dimensional time-independent Schr\"{o}dinger equation allows
the separation of variables not only in the cartesian system
$(x,y) \in I\!\!R^2$ but also in polar coordinates where it
degenerates to the P\"{o}schl-Teller problem in the quadruple-well
potential,
 \ben
\left (
-\frac{d^2}{d\,\varphi^2} + \frac{g}{\cos^2 2\varphi}
\right )\,\psi(\varphi)
=k^2\,\psi(\varphi), \ \ \ \ \ \ \ \varphi \in (-\pi,\pi).
\label{Pavel}
 \een
In the standard Hermitian setting the latter equation decays in
the four independent and exactly solvable eigenvalue problems with
$2\varphi \in (k\pi,k\pi+\pi)$ and $k = -2,-1,0, 1$, respectively.
The symmetry
 \ben
[H,{\cal R}] = 0
\label{syme}
 \een
of the Hamiltonian with respect to the shift ${\cal R}: \varphi
\to \varphi + \pi/2$ resembles the parity once we put ${\cal P} =
{\cal R}^2$.

Let us now consider a breaking of the Hermiticity of the
Hamiltonian. By a suitable complex deformation of the coordinate
line, $\varphi = \varphi(t)=t + i\,\varepsilon(t)$, $t  \in
(-\pi,\pi)$ we can avoid the barriers (i.e., poles
of the potential which lie at the integer multiples of $\pi/2$)
so that a tunneling takes place. Still, the energies need not become
complex after such a regularization of the potential
(cf. the three recent solvable examples in \cite{z}),
provided that we restrict our attention to the
complexifications which preserve the commutativity
 \be
[H,{\cal RT}] = 0.
\label{ysyme}
 \ee
As long as we have ${\cal P}^2 = {\cal R}^4=1$, our model is
not ${\cal PT}$ symmetric. Having a sample bound-state solution
$|\psi\rangle$ and the new symmetry (\ref{ysyme}) we
infer that the state ${\cal R}|\psi\rangle$ lies within the same
Hilbert space and satisfies Schr\"{o}dinger differential
equation at the identical energy. Superpositions
 \ben
|\psi^{[k,l,m,n]}\rangle=
(1+i{\cal R})^k
(1-i{\cal R})^l
(1+{\cal R})^m
(1-{\cal R})^n
|\psi\rangle
 \een
such that
 \ben
{\cal R} |\psi^{[0,1,1,1]}\rangle=i\, |\psi^{[0,1,1,1]}\rangle,\ \ \ \ \ \
{\cal R} |\psi^{[1,0,1,1]}\rangle=-i\, |\psi^{[1,0,1,1]}\rangle,
 \een
 \ben
{\cal R} |\psi^{[1,1,0,1]}\rangle=- |\psi^{[1,1,0,1]}\rangle,\ \ \ \ \ \
{\cal R} |\psi^{[1,1,1,0]}\rangle=+ |\psi^{[1,1,1,0]}\rangle.
 \een
can be expanded in a basis
$|n^{[k,l,m,n]}\rangle$ with $n = 0, 1, \ldots$ and with the
superscript which marks the symmetry. Due to the Schur's lemma,
the Hermitian Hamiltonian matrix becomes block-diagonal
and is partitioned accordingly.

Paralleling the two-by-two partitioning of ${\cal PT}$ symmetric
Hamiltonians, we now have a freedom of adding interactions
compatible with the four by four partitioning specified by the
four different ${\cal R}-$parities. Such complexifications
should obey any one of the conjugations $H=H^\ddagger$ as listed in eq.
(\ref{tabulka}). In the light of what has been said before, we
may expect {\it a priori} that at least a finite number of
energies $E_n$ remains real for a number of non-Hermitian interaction
terms.

\section{Recurrences and re-orderings of the basis}

One could construct the further conditions $H = H^\ddagger$
based on the partitioning $N$ by $N$ with $N=5$ etc. The
construction is recurrent in $N$. A key to its efficient
simplification exists and lies in a modification of the
projection technique. One has to re-order the basis states and
check how this changes the schemes of the type (\ref{tabulka}).
The result is unexpected since all the complicated multiply
partitioned solutions prove reducible to the single two by two
structure of eq. (\ref{SEdrd}) with the non-equal partitioning
dimensions in general. The ``generic", $(m+n)\times
(m+n)-$dimensional Schr\"{o}dinger operator reads
 \ben
\left (
\begin{array}{cc}
G-E\,I&-C\\
C^\dagger&
Z-E\,I
 \ea
\right )
\label{seseSEtre}
 \een
with, by assumption, ${\rm dim}\,G = m$, ${\rm dim}\,Z = n$
and $G=G^\ddagger$, $Z=Z^\ddagger$.
The inverse matrix exhibits the same structure,
 \ben
\left (
\begin{array}{cc}
G-E\,I&-C\\
C^\dagger&
Z-E\,I
 \ea
\right )^{-1}=
\left (
\begin{array}{cc}
G'(E)&-C'(E)\\
\left(C'\right)^\dagger(E)&
Z'(E)
 \ea
\right )\ .
\label{sesere}
 \een
In the mathematical induction step, the dimension
increases by one. With a new one-dimensional partition added on
the top,
 \ben
\left (
\begin{array}{ccc}
F-E\,I&\alpha\,A&
\beta\,B
\\
A^\dagger&
G-E\,I&-C\\
B^\dagger&
C^\dagger&
Z-E\,I
 \ea
\right )\
\label{peseSEtre}
 \een
the effective secular equation is one-dimensional,
 \ben
F- \left [
\begin{array}{cc}
( \alpha\,A&\beta\,B) \\
&
 \ea
\left (
\begin{array}{cc}
G'(E)&-C'(E)\\
\left(C'\right)^\dagger(E)&
Z'(E)
 \ea
\right )
\left (
\begin{array}{c}
A^\dagger\\
B^\dagger
 \ea
\right )
\right ]
 = E.
\label{pseFEdrd}
 \een
The effective Hamiltonian remains real if and only if
  \be
 \alpha = -\beta.
 \label{invv}
  \ee
This is the only condition required. We can very easily permute
the basis and re-derive all the three by three solutions
(\ref{tabl}) as well as all the four by four schemes in eq.
(\ref{tabulka}) etc. The latter case with $m+n=4$ is the first
one which gives either the square-shaped $A$ with $m=n=2$ or the
oblong blocks $A$ with dimensions $3=\max (m,n) > \min(m,n)=1$.

We may conclude that the recurrent picture is consistent. At any
partitioning $N$ by $N$ the number ${\cal K}$ of the
sub-partitions with the minus sign ($\alpha=-1, \ldots$) is not
arbitrary. Our construction admits one and two minus signs in the
respective two by two and three by three partitioned matrices.  At
the higher $N>3$ our choice becomes non-unique and we can opt for
the non-equivalent generalized ``weakly non-Hermitian" structures
of the Hamiltonian numbered by ${\cal K} = m \cdot n =
1\cdot(N-1)$ or $ 2\cdot(N-2)$ etc.  In each case a re-ordering of
the basis states transforms the Hamiltonian into the canonical two
by two structure
  \be
H = H^\ddagger=
\left (
\begin{array}{cc}
F&-A\\
A^\dagger&
G
 \ea
\right ),\ \ \ \ \ \ \
m={\rm dim}\,F,\ \  n={\rm dim}\,G\ .
\label{genre}
  \ee
with the negative sign attached to the $m \times n$ matrix
elements in the upper right submatrix of the Hamiltonian in
Schr\"{o}dinger eq. (\ref{SEdrd}).

Once we increase the number of partitions ${\cal M} =m+n$ of the
Hamiltonian matrix by one, the necessary and sufficient condition
(\ref{invv}) simply adds $m$ {\em or} $n$ blocks of minuses in the
upper line of the new partitioned matrix. In the former case we
can replace the upper-partition dimension $m$ by $m+1$ after we
permute the basis re-shuffling its topmost item to the bottom. In
the latter case the two by two partitioning is unchanged and we
replace $n$ by~$n+1$.

\section{Summary}

Generically, the models with non-Hermitian Hamiltonians $H \neq
H^\dagger$ possess the complex eigenvalues and make the evolution
non-unitary. This is the reason why their so called ${\cal PT}$
symmetric special cases can be considered as an appealing
alternative to their Hermitian predecessors. We have seen that
there exists a formal connection between the Hermitian and ${\cal
PT}$ symmetric form of $H$. It is based on the similarity between
their non-linear (so called effective) reductions constructed by the
Feshbach's projection method~\cite{Feshbach}.

We described in detail this intimate relationship (i.e.,
sign-difference) between the respective Hamiltonians as
well as effective Hamiltonians (the latter proved Hermitian in
both these cases). We emphasized that even in the
non-Hermitian, ${\cal PT}$ symmetric case the Schr\"{o}dinger
equation generated many (if not all) energies for a ``very broad"
range of the Hermiticity-violating components of the interaction.

The latter observation inspired an immediate generalization of the
${\cal PT}$ symmetry to a more general property (\ref{genre}). It
is characterized by the partitioning dimensions $m,n$ and
degenerates to the Hermiticity at $m=0$ or $n=0$ and to the ${\cal
PT}$ symmetry at $m=n\neq 0$. In all the non-Hermitian cases with
$m>0$ and $n>0$ the left eigenvectors are
different from the right ones. The validity of the equation
 \ben
 \left (
 \begin{array}{cc}
 G-E\,I&-C\\ C^\dagger& Z-E\,I
 \ea
 \right )
 \left (
 \begin{array}{c}
 \vec{g}\\
 \vec{z}
 \ea
 \right ) = 0
 \een
implies that
 \ben
 \left [
 \begin{array}{cc}
 (\vec{g}, -\vec{z})\\
 &
 \ea \!\!\!
 \left (
 \begin{array}{cc}
 G-E\,I&-C\\ C^\dagger& Z-E\,I
 \ea
 \right )
 \right ]=0.
  \label{stre}
 \een
The related ``natural" normalization remains indefinite in its
sign,
 \be
 \sum_{j=0}^m\, \left (g_j
 \right )^2
 -
 \sum_{k=0}^m\, \left (z_k
 \right )^2 = \pm 1.
 \label{selfoverlap}
 \ee
This can be interpreted as a result of an overlap
between the right eigenvector $|\psi \rangle$ and its
new
conjugate $\langle \langle \psi |=\langle \psi| {\cal Q}$.
The ``metric" ${\cal Q}$ is
a unit matrix with the last $m$
diagonal elements replaced by $-1$. In the
${\cal PT}$ symmetric special case where $m=n$ this operator
coincides with the parity ${\cal P}$ .

The innovation of the bra vector leads to the
modified inner product. It exhibits the
(pseudo-)orthogonality feature
 \be
 \langle \langle \psi_j | \psi_k \rangle =
 \langle \psi_j |{\cal Q}| \psi_k \rangle = \pm \delta_{jk}
 \label{product}
 \ee
when computed between the two different eigenstates of $H$. The
alternative inner product has been used
in many $m=n$ studies of the perturbations of Hermitian
Hamiltonians (cf. ref. \cite{Simon} and, especially,
Corollary II.7.6 there) as well as in the early stages of
development of the elementary ${\cal PT}$ symmetric models (cf.
\cite{alvarez} and eq. (14) there, or the text
after eq. (6) in ref. \cite{mytri}).

In the present context, the use of the $m \neq n$ ``metric"
${\cal Q}$ leads
to a natural generalization of the concept of the Hilbert
space \cite{ioch}. The self-overlaps
(\ref{selfoverlap}) [or (\ref{product}) at $j=k$] can be
understood as a pseudo-norm in a space where the time-evolution is
pseudo-unitary \cite{pseudo}. This opens new
perspectives and questions (e.g., about the possible physical
interpretation of the wavefunctions) shared by our present
generalized formalism with its increasingly popular ${\cal PT}$
symmetric predecessor~\cite{pseudob}.

\section*{Acknowledgement}

Supported by GA AS (Czech Republic), grant Nr. A 104 8004.

\end{document}